# THE ARCHAIC UNIVERSE: BIG BANG , COSMOLOGICAL TERM AND THE QUANTUM ORIGIN OF TIME IN PROJECTIVE COSMOLOGY

**Ignazio Licata**

*Institute for Scientific Methodology*

*Palermo University*

*Italy*

*Ignazio.licata@ejtp.info*

**Leonardo Chiatti**

*AUSL VT  Medical Physics Laboratory*

*Via Enrico Fermi 15, 01100 Viterbo*

*Italy*

**SUMMARY**

This article proposes some cosmological reflections at the qualitative and conjectural level, suggested by the Fantappié-Arcidiacono projective relativity theory. The difference will firstly be discussed between two types of singularity in this theory: geometric (de Sitter horizon) and physical (big bang, big crunch). The reasons for the existence of geometric singularities are deeply rooted in the principle of inertia and in the principle of relativity, while physical singularities are associated with the creation or destruction of matter.

In this framework, quantum mechanics is introduced through a particular interpretation of Bohm's holomovement. Finally, a possible mechanism is discussed for the genesis of the cosmological term. No form of inflation appears in the scenario described.

## 1. INTRODUCTION

In a recent work, Kutschera and Dyrda [1] highlighted a rather intriguing coincidence. They start from the equation giving the age of the Universe $t_0$ in every relativistic cosmological model:

$$H_0 t_0 = \int_0^\infty \frac{dz}{(1+z)E(z)} \;; \tag{1.1}$$

where $H_0$ is Hubble's constant and:

$$E(z) = \left[\Omega_M(1+z)^3 + \Omega_R(1+z)^2 + \Omega_\Lambda\right]^{1/2} . \tag{1.2}$$

Here, $\Omega$ is the matter-energy density to critical density ratio, and $\Omega_M$, $\Omega_R$, $\Omega_\Lambda$ are the contributions to $\Omega$ relating to matter, to curvature and to the cosmological constant, respectively; the $\Omega_M + \Omega_R + \Omega_\Lambda = 1$ constraint applies. For a flat model, we have $\Omega_R = 0$ and, by imposing the $H_0 t_0 = 1$ relation valid for Milne cosmology, we obtain, by means of a simple calculation: $\Omega_M = 0.26$, $\Omega_\Lambda = 0.74$. These are precisely the values of the mass parameters derived from the $\Lambda$CDM concordance cosmological model. In other words, the relation $H_0 t_0 = 1$, which is exactly true for Milne kinematic relativity formulated for an empty Universe, remains exactly true in the $\Lambda$CDM model where it leads to the same results confirmed by WMAP. In addition, it has been observed [2] that, even though data relating to Ia supernovae are better reproduced by the $\Lambda$CDM model, agreement with the Milne model stays good up to a distance of 8 billion light years.

We believe that these strange coincidences can derive from the fact that, contrary to what is assumed in general relativity, the $t_0$ parameter is a fundamental constant independent of the distribution of matter and energy. This article puts forward, solely at a qualitative and descriptive level, some conjectures on the expansion of the Universe, based on a theory of relativity where $t_0$ actually appears as a fundamental constant; that is, the Fantappié-Arcidiacono projective relativity.

This underrated theory [see ref. 3 for a review], several times rediscovered independently [4, 5], is indeed the natural heir to Milne's kinematic relativity. This theory, too, does not appear to give rise to glaring contradictions with observational data, though further quantitative studies seem appropriate and necessary [6]. In any case, some preliminary specifications on the nature of the singularities are required. Even a preliminary investigation shows [6] that, contrary to Arcidiacono's arguments [3], the past and future De Sitter horizon cannot be identified with the singularities commonly known as the big bang and the big crunch, respectively. One must therefore distinguish between two different classes of singularities: geometric singularities (de Sitter horizon) and physical ones (big bang, big crunch). The origins of these classes are different and request separated discussions.

Firstly, we propose an introduction to Fantappié-Arcidiacono relativity, which attempts to highlight how the de Sitter horizon, and the light cone as well, are induced by fundamental principles of classical physics, such as the principle of inertia and the principle of relativity. Successively, quantum phenomena are introduced in this framework through a particular implementation of Bohm's "holomovement" [7]; this part is a development of ideas originally outlined by Arcidiacono between 1960 and 1990 ("hyperspherical" archaic universe).

A possible mechanism is then proposed for the origin of the cosmological constant, based on considerations originally developed by Milne (double time scale). Since, according to this hypothesis, the cosmological constant not undergoes to dynamical evolution, inflation is not possible and an alternative to it is needed. This leads to the hypothesis of an initial isotropic (physical) singularity, whose homogeneity is induced by the self-consistency of holomovement.

## 2. PRE-QUANTUM CONSIDERATIONS

Let us represent the motion of a material point using a system of three spatial coordinates $x$ and a time coordinate $t$; the elementary motion will thus be defined by the displacement $(x, t) \rightarrow (x + dx, t + dt)$ or, alternatively, by the quantities $(x, t)$, $v = dx/dt$ and $dt$.

In ($x$, $t$) the vector $v$ undergoes to a change due either to the bodies and fields acting on the material point in ($x$, $t$), or merely to the choice of coordinates (e.g. centrifugal effects in a rotating frame). One can therefore write, without losing in generality:

$$dv' = f_1(x', v', \tau)d\tau + f_2 \qquad (2.1)$$

where $v'$, $x'$, $\tau$ are defined starting from $x$, $v$, $t$ as will be specified below, and $dx' = v'd\tau$. The function $f_1$ describes the action on $v'$ of the bodies and fields present in ($x$, $t$) while $f_2$ (function of $x$, $t$, $dx$, $dt$) is a contribution to $dv'$ derived from the choice of coordinates. If the identity $f_2 = 0$ holds, the reference frame where $x$, $t$, $dx$, $dt$ are evaluated is called inertial; in this frame $v'$ is constant with respect to $\tau$ when $f_1 = 0$, i.e. when no interaction is present.

If we postulate the existence of at least one inertial frame (1$^{st}$ principle of dynamics), the law (2$^{nd}$ principle of dynamics)

$$dv'/d\tau = f_1(x', v', \tau) \qquad (2.2)$$

must apply within it.

The vector $f_1$ is the force per unit mass acting on the material point. If the material point is free, $f_1 = 0$ and therefore $v'$ is constant in $\tau$; consequently, the function $x'(\tau)$ is linear. This result is naturally valid for every system of coordinates ($x$, $t$) provided that it is inertial. The passage from one inertial system of coordinates to another thus has the effect of transforming a linear trajectory $x' = x'(\tau)$ into a new linear trajectory. This transformation of coordinates on the space $\{x, t\}$ must therefore induce a *linear* transformation of coordinates on the space $\{x', \tau\}$.

Now, the most general linear transformation of coordinates is the projectivity, so that, given a reference frame in the space $\{x', \tau\}$ associated with an inertial system of coordinates ($x$, $t$), the generic projectivity changes this reference into a new reference which is also associated with an inertial system of coordinates.

Two elementary motions relating to the same material point or to different material points can occur in two different locations in the space $\{x', \tau\}$; in this case, a causal connection between them is possible only if a timelike universe line exists that joins such locations, and therefore a proper time interval T' (counted along this line) that separates them. In the case that the two events (relating to different material points) coincide in $x'$ and $\tau$, they can be distinguished by the value of $v'$; in this latter case, they are only separated by a relative velocity vector of modulus Δ' and there is no impediment to the possibility of a causal link between them.

The crucial observation at this point is that the only invariant of the projectivity consists of the cross-ratio. Therefore, in order to express an invariant law for transformations connecting inertial reference frames (as is the 2$^{nd}$ principle of dynamics) through the parameters T' e Δ', these parameters must be represented by cross-ratios. It is now possible to define the passage from the original coordinates $x$, $t$, $v$ to the parameters $x'$, $v'$, $\tau$ to be substituted in the 2$^{nd}$ principle.

Let us first of all postulate that the original space $\{x\}$ is real, Euclidean and three-dimensional and that the distance $d$ from the origin is expressed by Pythagoras' theorem:

$$d^2 = x_1^2 + x_2^2 + x_3^2. \qquad (2.3)$$

Let us postulate that the original space $\{v\}$ is real, Euclidean and three-dimensional and that the distance Δ from the origin is expressed by Pythagoras' theorem:

$$\Delta^2 = v_1^2 + v_2^2 + v_3^2. \qquad (2.4)$$

Let us postulate that the original time $t$ is a real, one-dimensional, Euclidean space with distance from the origin expressed by T = $|t|$.

Let us consider the universe line on which the two events we discussed above are placed and let Q, P be the representative points of these events. As we have to express the distance between these two points in terms of a cross-ratio that is independent of other "real" points, two extreme "ideal" points on the line must exist which we shall indicate as O and S [Fig. 1]. The cross-ratio will therefore be expressed by the relation:

$$(QPOS) = (OQ/PO)(PS/QS) \ .$$

QP is the proper time interval T between the two events, measured in the space $\{x, t\}$ and, if we choose Q as the origin of time, then T = $t$. OQ is the interval between the event Q and the beginning of time $t$, while SQ is the interval between Q and the end of time $t$. The cross-ratio then becomes:

$$(QPOS) = (OQ/SQ)(PS/PO) = (OQ/SQ)[(SQ - t)/(OQ + t)] \ .$$

By rigidly translating the segment PQ on the line OS the distance (i.e., the time interval $\tau$) T' between the two events Q and P must remain unchanged, because we are attempting to define a concept of distance which is independent of position. Therefore T' cannot depend explicitly on Q but only on $t$. It follows that SQ and OQ must be independent of Q and equal to two constants $t_0'$, $t_0''$. A second observation is that the value of T' cannot depend on which of the two extremes O, S we choose as the beginning (end) of time $t$. It follows that the two constants $t_0'$, $t_0''$ must be equal. From here onwards, let $t_0' = t_0'' = t_0$. Hence:

$$(QPOS) = [(t_0 - t)/(t_0 + t)] \ .$$

For the purpose of having an additive definition of distance, let:

$$T' = k \, log(QPOS) = k \, log \, [(t_0 - t)/(t_0 + t)] \ .$$

The value of $k$ can be found by taking into consideration the limit of this expression for $t_0 \to \infty$. We have:

$$T' = -k \, log \, [(t_0 + t)/(t_0 - t)] \to$$

$$-k \, log \, [(1 + t/t_0)(1 + t/t_0)] =$$

$$-2k \, log \, (1 + t/t_0) \approx -2k \, log \, exp(t/t_0) = -2k \, t/t_0 \ .$$

Let $k = -(t_0/2)$; we have T' = $t$ and, placing the origin of proper time in point Q also in the space $\{x', \tau\}$, we have T' = $\tau$, so that in conclusion:

$$\tau = (t_0/2) \, log \, [(t_0 + t)/(t_0 - t)] \ . \qquad (2.5)$$

If, on the other hand, the two events coincide on the space $\{x', \tau\}$ but are separated by a relative velocity with modulus $\Delta'$, then the line that joins them must be imagined on the one-dimensional space of the relative velocity vector moduli. In this case QP = $\Delta'$, while OQ and SQ are the differences between the current velocity of the material point (represented, say, in Q) and the two limit velocities in opposite directions. Again, the distance $\Delta$ must not depend explicitly on Q nor on the sign adopted for the velocities, thus OQ = SQ = $c$, where $c$ is a constant independent of Q. Thus:

$$\Delta = k\,log(\text{QPOS}) = k\,log\,[(c - \Delta')/(c + \Delta')] = -\,k\,log\,[(c + \Delta')/(c - \Delta')]\ ;$$

$$d\Delta = -2ck\,d\Delta'\,/\,(c^2 - \Delta'^2)\ .$$

Since Q is chosen as the origin of the velocity space, $\Delta' = v'$ while $\Delta = v$. In strict conformity with the case of the distances, let $k = -c/2$; we therefore obtain:

$$dv = dv'\,/\,(1 - v'^2/c^2)\ . \qquad (2.6)$$

We notice that if the space is isotropic (which is certainly true if it is empty), the maximal velocity $c$ cannot depend on the direction of the relative velocity vector of the elementary motion events Q, P. In other words, the limit speed $c$ must be the same in all spatial directions. The meaning of expressions (2.5), (2.6) will be discussed in the following section.

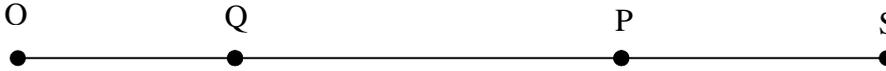

Fig.1

## 3. THE PRINCIPLE OF RELATIVITY

In relation (2.6) we have $v' \in (-c, +c)$ for any inertial reference; it follows that the region of modified spacetime $\{x', \tau\}$ accessible by Q is that which falls within the cone $|v'| = |dx'/d\tau| \leq c$. If we momentarily forget relation (2.5), this result means that the transformations between inertial references must preserve the quadric:

$$c^2\tau^2 - x_1'^2 - x_2'^2 - x_3'^2 = 0\ . \qquad (3.1)$$

As is known, their set constitutes the Poincaré group, which admits the general invariant:

$$c^2\tau^2 - x_1'^2 - x_2'^2 - x_3'^2 = c^2 t^2. \qquad (3.2)$$

These arguments must now be extended to include equation (2.5). The time $t$ that appears in equation (3.2) is that measured by a local clock in a state of rest (proper time) and therefore coincides with the time $t$ of equation (2.5). Analysis of equation (2.5) shows that $t$ is included between the extreme values $-t_0$ and $+t_0$; we must therefore have $t^2 \leq t_0^2$ or, from (3.2):

$$c^2\tau^2 - x_1'^2 - x_2'^2 - x_3'^2 - r^2 \leq 0,$$

where $r = c\, t_0$. To take into account equation (2.5), the passage is therefore required to new group of transformations, more extensive than the Poincaré group, which leaves unchanged the quadric:

$$c^2\tau^2 - x_1'^2 - x_2'^2 - x_3'^2 - r^2 = 0. \qquad (3.3)$$

As is known, this is the de Sitter-Fantappié group [3], admitting the Poincaré group as its limit for $r \to \infty$.

One may wonder about the reason for the transformation of coordinates $(\mathbf{x}, \mathbf{v}, t) \to (\mathbf{x'}, \mathbf{v'}, \tau)$. It is the following: the motion equation (2.2) must be valid in every inertial reference system, and must therefore stay valid under transformations that give rise to the passage from one inertial reference system to another. If the "native" Galilean coordinates $(\mathbf{x}, \mathbf{v}, t)$ were to be adopted, we would notice that the left-hand member of equation (2.2) actually preserves its form under the relevant group of transformations (Galileo group), but this is not generally true for the right-hand member. For example, electromagnetic interactions do not preserve their shape under Galileo transformations. The passage to the new coordinates (no longer Euclidean-Galilean) $\mathbf{x'}$, $\tau$ and to the new impulse expression transforms the Galileo group into the Poincaré group, thus allowing the invariance group of the left-hand member of equation (2.2) to be made homogeneous with the invariance group of the right-hand member. Now both members of this equation preserve their form under the same group of transformations, which is the Poincaré group. This ensures that equation (2.2) is actually valid in every inertial system.

If only one force field existed in the physical world, it would always be possible in principle to perform a transformation of coordinates of the type mentioned above, so that the mechanics would be made to conform with the covariance group of that field's equations. This would not involve any new physical principle, but only a useful convention. Yet, in the physical world several distinct interaction fields exist and the surprising experimental result is that the transformation of coordinates required to make equation (2.2) valid in every inertial reference is the same regardless of the fields appearing in the right-hand member. This fact, which is formally expressed by the foregoing reasoning, is a physical principle of the greatest importance: the principle of relativity. The de Sitter-Fantappié group is the covariance group of all physical laws.

The theory of relativity based on the assumption that physical laws are invariant with respect to the Poincaré group is special relativity (SR), while that which assumes that the laws are invariant with respect to the de Sitter-Fantappié group is projective special relativity (PSR) [3]. SR is the limit case of PSR for $r \to \infty$.

We owe this manner of introducing the relativity principle to Tyapkin [8]. It has the merit of doing away with the mystery of the constancy of $c$ and $t_0$ for all observers; indeed, there is nothing mysterious about this constancy. A reference frame is defined as inertial when the impulse and the energy of a free material point are conserved within it, so that the inertial reference definition is closely connected to that of impulse and of kinetic energy. The form of the impulse which appears in the expression of the second principle of dynamics depends, in turn, on how an external interaction modifies the motion of the body; it is a description of the body's inertial properties with respect to its interaction with external fields. An inertial system is therefore defined as such by the properties of the interactions, not by space and time properties. Different approximations in the study of interactions between bodies and fields give rise, therefore, to different descriptions of the inertia of bodies and, consequently, to different concepts of an inertial system, i.e. to different theories of relativity. But space and time remain basically Euclidean and Galilean (eq. 2.3, 2.4).

Consequently, $c$ and $t_0$ are properties of interactions, not of space or time; and, indeed, they are connected with the limit velocity of propagation of the influence generated by an interaction and with the maximum duration of this influence. The passage from the Euclidean-Galilean to the PSR or SR chronotope corresponds to the choice of a particular description of inertia and its purpose is to replace Galilean calculation of the coordinates with a new calculation that makes the two members of equation (2.2) covariant with respect to the same group of transformations. The coordinates redefined in this way constitute a chronotope, which is no longer Galilean-Euclidean, incorporating $c$ and $t_0$ as its geometric features. The metric element of this chronotope is an invariant of the symmetry group of equation (2.2) and by adopting its coordinates the parameters $c$ and $t_0$ become constants that are the same for all observers. It must be noted that while the choice of the calculation of coordinates is totally conventional in itself, the possibility of choosing a calculation that satisfies covariance requirements for all fields appearing in $f_1$ is an objective property of nature (principle of relativity). From this point of view, Lorentz and Poincaré vision of relativity was therefore correct; they persisted, however, in identifying the fundamental Euclidean-Galilean description with a useless and impossible etheric material "fluid".

The approach adopted here also makes the passage from the "special" theories introduced (SR and PSR) to the corresponding "general" theories (GR and PGR) [3] which include gravitation more comprehensible. It is plausible that, once a given "special" theory (SR or PSR) has been established, the definition of the local class of inertial systems changes from the pointevent P to an infinitely close pointevent P'. This means stating that the value of the free material point impulse changes in passing from P to P'; i.e., the body undergoes "spontaneous" acceleration. If this acceleration is independent of the mass of the body (Galilei principle) then it can be incorporated in a suitable "deformation" of the chronotope, thus obtaining the corresponding general theory (GR = general relativity or PGR = projective general relativity, respectively). With this passage a calculation of coordinates is adopted which is dependent upon the position P, in such a way as to incorporate the effects of the acting field (gravitation) in the free motion. Yet the spacetime reference structure is and remains Euclidean-Galilean: true spacetime does not curve and does not undergo any distortion due to the matter present within it.

Some remarks:

1) One can observe that the existence of limiting surfaces and the principle of inertia are closely interlinked facts. In informal terms, we request a spacetime where the universe lines of free material points are straight. Thus every transformation of spacetime that changes lines into lines leaves the dynamics unchanged. The more general collineation, on the other hand, is a projectivity, so that these transformations are actually projective transformations. Projective transformations can convert "real" points into "ideal" points, and the existence of limiting surfaces such as the quadric described by equation (3.3) follows from this.

2) A similar argument can be applied to velocity vector space in SR. The more general transformation of a vector into a vector is again a projectivity. The limiting surface in this space will be physically associated with the existence of a limit propagation velocity. Thus, the appearance of such velocity is in no way magic or surprising.

## 4. HYPERSPHERIC PRESPACE

Let us consider, in 5-dimensional real Euclidean space, the surface represented by the equation:

$$r^2 = w^2 - c^2 t^2 + x^2 + y^2 + z^2 ,  \qquad (4.1)$$

where $r=ct_0$ is a constant. Using the origin as the centre of projection, let us project the points of this surface onto the plane tangent at a generic point Q belonging to it. This plane (Castelnuovo chronotope) contains both a light cone with origin in Q and the two limiting surfaces (past and future with respect to Q) whose equation is given by (3.3). By applying the substitution (Wick rotation) $t \to jt$, (4.1) becomes the equation of the 5-spherical surface with radius $r$ having its centre at the origin. Arcidiacono showed that the de Sitter-Fantappié group is isomorphic to the group of rotations of this surface around its centre [3]. These rotations change, if the tangent plane is kept fixed, or the point Q or the reference system associated with it. In the first case, there is a passage from an observation pointevent Q to a new observation pointevent Q'; in the second case, a rotation or a boost is effected on the reference having its origin in Q. In general, a combination of the two circumstances will occur.

The tangent plane thus constitutes the "private chronotope" where the observer placed in Q coordinates events, while (4.1) defines the "public chronotope" formed by all possible observation events Q. The fifth coordinate is not needed for coordination of events and is therefore not "observable". It is merely a calculation stratagem for linking observations made by different observers. It is real solely in the sense of the intrinsic geometry of Castelnuovo spacetime and not in the extrinsic sense of an actual immersion in the spacetime of a pentadimensional space[1].

One immediately realises that the theory, contrary to Arcidiacono's belief [3, 9], cannot be generalised to N > 5 dimensions. In this case, the hypersphere surface would become (N-1)-dimensional and the plane tangent to it at a given point would also be (N-1)-dimensional. So if we continued to interpret such a plane as the private spacetime in which an observer associated with the tangent point coordinates events, we would have the appearance of N-1-4 = N-5 *physical* coordinates in addition to the spacetime ones. Such coordinates, uncompacted, ought to be observable and are not.

We can therefore draw the following conclusion: assuming the first two principles of dynamics and the relativity principle, and assuming an original Galilean-Euclidean ambient space, it follows that an inertial observer coordinates events in a Castelnuovo chronotope. This chronotope can be derived from a 5-dimensional hypersphere through the following succession of commutable operations:

1) geodesic projection on the 4-dimensional tangent plane passing through the observer (the point of projection is the centre of the hypersphere),

2) Wick rotation $t \to jt$.

Clearly, this reasoning can be inverted: given the 5-sphere as prespace, by applying the two operations described in arbitrary order we obtain the Castelnuovo chronotope, on which a class of inertial observers is thus defined. This result applies for PSR but, taking into account the statements made in the previous section, there would seem to be no problems with extending it to the PGR domain. A generic inertial observer will then simply be constructed by connecting geodesic segments, each corresponding to a *local* inertial observer. The manifold to be projected onto the tangent plane at its generic point will no longer be (4.1), but a more general solution of the PGR gravitational equations.

---

[1] Each point, whether real or ideal, of the Castelnuovo 4-dimensional plane becomes, through the introduction of a fifth homogeneous coordinate, a line in 5-dimensional Euclidean space [3].

PSR applies for an empty space without matter [6] and therefore it cannot describe any matter-creating phase (big bang) nor any destructive phase (big crunch). Nevertheless, the passage to PGR presupposes the presence of matter; thus, these phases can exist. It is plausible that the big bang and the possible big crunch are linked to isotropic singularities of the PGR metric (as we will see ahead). If the cosmological principle is valid, operation 2) can be generalised in $t \rightarrow jt'$, where $t'$ will be an appropriate cosmic time defined on the basis of the PGR metric. This cosmic time will be counted starting from the big bang singularity and will end at the possible big crunch singularity. If the state of matter along the big bang singularity is homogeneous, the various cosmic time lines are equivalent, since they correspond to observers who - unless statistical fluctuations - will see the Universe evolve in the same way. The homogeneity of the early Universe is thus closely linked to the cosmological principle.

## 5. ENTER QUANTUM MECHANICS

Let us now see how to introduce the most basic concepts of quantum mechanics (QM) into the framework outlined in the previous section. If we refer to the wave formulation of QM, a certain wavefunction is associated with each elementary particle. Let us assume that prespace is constituted by the 5-dimensional hypersphere in the Euclidean space $E_5$:

$$y_1^2 + y_2^2 + y_3^2 + y_0^2 + y_5^2 = r^2.$$

Let us consider an arbitrary point on it to which we shall assign, without any loss of generality, the coordinates $y_i = 0$ ($i = 0, 1, 2, 3$), $y_5 = r$. Let us consider, in a sufficiently small neighbourhood around this point, the single particle wavefunctions of the form $exp(\lambda^i y_i)$ [this reasoning can easily be extended to many particle wavefunctions of the form $exp(\lambda^{ik} y_{ik})$, where $k$ is the particle index]. These functions are evanescent or emergent, if the $\lambda$ and $y$ parameters are real.
We shall postulate that the *creation* of a particle on spacetime takes place through the following succession of commutative transformations of wavefunction coordinates.

I - 1$^{st}$ passage.
It consists of a Wick rotation $\lambda^i \rightarrow j\lambda^i$, $y_0 \rightarrow -jy_0'$, where $j$ is the square root of -1 and $i = 1, 2, 3$. As a result of this operation, the evanescent/emerging waves become harmonic. The effect is similar to that experencied by a wave when it pass through a barrier of finite height and extension in the tunnel effect.

II - 2$^{nd}$ passage.
The thus obtained chronotope is geodesically projected onto the plane tangent to it at the image point of the original $y_i = 0$ ($i = 0, 1, 2, 3$), $y_5 = r$. The coordinates are thus changed into the "actual" physical coordinates $x^k$ ($k = 1, 2, 3$), $t$. The variable $t$ does not indicate an arbitrary proper time, but a cosmic time. The two passages can therefore be effected only if a coordinate $y_0' \rightarrow t$ exists, i.e. if the cosmological principle is valid. We shall come back to this point later.

After these passages[2], the waves are usual plane waves of the form $exp(jEt - j\mathbf{p} \bullet \mathbf{x})$. A linear superposition of these waves:

$$\psi(x_i, t) = \int dE\, dp_i\, \alpha(E, p) \exp(jEt - j\vec{p} \bullet \vec{x})$$

constitutes the generic possible wavefunction for a real particle with mass $M$, if $E^2 = p^2 + M^2$. If the functions $\alpha$ are scalar, then $\psi$ locally satisfies the Klein Gordon equation, as can be immediately verified. We can also see that if the functions $\alpha$ are quadrispinor, the $\psi$ satisfies the Dirac equation, and so on. The $\psi(x_i, t)$ thus obtained is the wavefunction of a particle created at the cosmic time instant $t$. After its creation, the wavefunction evolves as a result of the interactions, in conformity with an equation of the form $H\psi = j\partial_t \psi$. The initial condition of this equation is constituted by $\psi(x_i, t)$.

We shall then postulate that the *annihilation* of a particle is associated with the succession of commutative transformations constituted by the inverses of the two transformations described; these however must be understood as being applied not to the *initial* wavefunction $\psi$ of the particle, but to the *final* wavefunction following the reduction of the wave packet. The processes thus described will therefore constitute Penrose nonunitary "**R**" processes, and one can see that the pair of events constituted by the creation of a particle and by its subsequent annihilation is a "similitude metamorphosis" according to the Bohm definition [7]. The creation corresponds to the unfolding, in the explicate order (spacetime), of the information contained within the implicate order (prespace); annihilation corresponds to the inverse process of enfolding. The complex of all these processes (holomovement) connects spacetime to prespace, which is *per se* atemporal; we are thus dealing with an evolution which is no longer just diachronic but which also has sinchronic features[3]. A terminological specification is thus called for: in usual diachronic cosmology, the adjective "early" is used to indicate the state of the Universe immediately after the big bang; in order to avoid any confusion, it is advisable that in a cosmology having a prespace, the state of the Universe before the appearance of the time axis (clearly "before" must be understood here in a logical or ontological, not chronological sense) be indicated with the adjective "archaic". Our "archaic Universe" is the 5-sphere.

*The postulated transformations act on the initial and final particle wavefunction coordinates, not on "empty" prespace coordinates. The latter remains unchanged.* On the other hand, one must remember that the value of a particle wavefunction, at a certain instant and in a certain position, is linked to the probability that that particle may be localized at that instant in that position. This partial or complete localization can only be carried out by means of an interaction with other particles; during this interaction, the previous wavefunction will be annihilated and a new one, associated with the new particles produced, will be created. In this sense, therefore, spacetime is simply the union set of all possible interaction regions or the domain set of all possible interaction vertices. It only actually exists to the extent that it is materialised in *actual* interactions.

In the course of a creation the 3-sphere with equation $y_0$=constant, included within the archaic 5-sphere, turns into a constant-time hyperboloid in the Castelnuovo chronotope (or into a surface with constant cosmic time in the relevant PGR generalisation). This fact distinguishes a given spatial direction on the 5-sphere as a time direction and leads to the emergence of the physical time. At the same time the chronotope metric arises with the related system of timelike geodesics that can be associated with inertial observers. It must be observed that the emergence of time, associated with Wick rotation, is an internal fact of wavefunctions and has no relationship with "external" geometry, which remains that of prespace. Time, so to speak, emerges as a fact that is internal to matter and is not a sort of pre-existing arena where phenomena take place.

---

[2] Note that by contracting the Castelnuovo chronotope by a factor $N_D = 10^{40}$ (Dirac's number) we have $t_0 \to \tau_0 \approx e^2/mc^3 = \rho/c$, where $\rho$ is the classical radius of the electron. A mass quantum $\eta c/\rho \approx 70$ MeV, as shown by Nambu [10] and McGregor [11], is recurrent in hadronic and leptonic masses. This correspondence does not exist in SR, because in SR $t_0$ is infinite.

[3] Some particles will be created when other particles will be annihilated, so that creations and annihilations will mutually imply each other. Thus, the diachronic dimension (spacetime) will dialogue with the synchronic dimension (5-sphere) through the annihilations, inducing new creations. Conversely, the latter will dialogue with the former through creations, inducing annihilations.

Wavefunctions on spacetime satisfy wave equations in which D'Alembert operators appear (each component of a spinor of any order satisfies the Klein-Gordon equation). Such operators, after annihilation or before creation, become 5-dimensional Laplacians and the wave equations become *static* 5-dimensional Laplace equations. Thus there is no sense in speaking of waves diverging from one point of the hypersphere that converge in another point, and therefore of indissoluble "syntropic-entropic" characteristics, as did Arcidiacono [12]: there is actually no wave propagation on the hypersphere. After the Wick rotation, the hypersphere becomes a twofold hyperboloid: the time becomes open, the future is distinguished from the past and is joined to it only in the present. Thus there is no longer any possibility of travelling through time loops[4].

On 5-spheric atemporal prespace, all and only initial (final) wavefunctions are mapped that correspond to *actual* creation/annihilation events of sets of particles which have occurred, are occurring or will occur in the history of the Universe. From this point of view elementary quantum processes are deterministic; however, as the information represented on the 5-sphere is not directly accessible except in the form of interactions on spacetime, there is scope for quantum randomness at the operational level.

## 6. THE BIG BANG AND THE BIG CRUNCH

If the cosmological principle is assumed valid, then a cosmic time exists at every instant of which all observers see the Universe in the same condition. As is well known, the dependence of the scale of distances on cosmic time derives from this postulate in a completely general fashion, with singularities corresponding to the big bang and the big crunch. In the terms expressed in the previous section, this means that particle creation/destruction events associated with elementary interactions occur within a limited $[0, t_{max}]$ cosmic time interval. Thus, on the chronotope that generalises the Castelnuovo one, we have the actual creation of matter (big bang) at time 0 and the possible actual destruction of matter (big crunch) at time $t_{max}$. The big bang is $t_0$ away from the past limiting surface identified by (3.3). The big crunch is $t_0$ away from the future limiting surface identified by (3.3) [3].

A generic inertial observer, by moving towards the past (future) along his universe line, at the end intersects the hypersurface $t = 0$ [$t = t_{max}$] where this line comes to an end. Before $t = 0$ (pre-big bang era) or after $t_{max}$ (post-big crunch era) the wavefunctions do not undergo the transformation (or "metamorphosis") described in the previous section. The density of matter and probably also the cosmological constant (please see the following section) become null, so that the de Sitter-Castelnuovo PSR metric holds. This metric has a Euclidean spatial section and null curvature. The answer to the naïve question "what was the world like before the big bang" can therefore be either in synchronic terms referring to the always existing 5-spheric archaic Universe, or in the usual diachronic terms referring to the empty de Sitter-Castelnuovo chronotope. The latter reply constitutes the essence of the Hartle-Hawking solution, as several works have already pointed out[5] [13, 14, 15].

---

[4] We might say that at its creation an elementary particle is "placed in motion" in time; at its annihilation this "race in time" comes to a stop.

[5] One might object that in the Hartle-Hawking solution we are dealing with a superposition of different metric states, while in this case the PGR metric is univocally defined. One can nevertheless speculate that the true fundamental theory of interaction between a gravitational field and matter is not metric, the metric description of gravitation being only an

# 7. CONJECTURE ON THE ORIGIN OF THE COSMOLOGICAL TERM

In a recent work [6] it has been shown that the PGR cosmic time $t$ and the $\tau$ one of usual Fridman relativistic cosmology are linked by the Milne scale transformation:

$$\tau = t_0 + t_0 \ln\left(\frac{t}{t_0}\right). \tag{7.1}$$

The variable $t$ is confined to values included between 0 and $t_0$, so that the $\tau$ domain is $(-\infty, t_0]$. It derives from (7.1) that free, rectilinear, uniform motion in time $t$, with equation $x = Kt$, in usual relativistic cosmology corresponds to an equation:

$$x = Kt_0 \exp\left(\frac{\tau - t_0}{t_0}\right) \tag{7.2}$$

whose second derivative with respect to $\tau$ is:

$$\frac{d^2 x}{d\tau^2} = \frac{K}{t_0} \exp\left(\frac{\tau - t_0}{t_0}\right) = \frac{x}{t_0^2}. \tag{7.3}$$

This expression coincides, owing to (2.2), with the force per unit mass experienced by a material point (which would be free in PGR) in a cosmological model based on usual general relativity.

In the usual Newtonian cosmology with a cosmological term $\lambda$, the material point placed at the distance $x$ from observer O is subjected to a force per unit mass composed of two terms [16]: attraction by the mass contained within the sphere with its centre in O and radius $x$ and the repulsive term $(1/3)\lambda x$. By choosing the $x$ axis along the direction of the $x$ coordinate of (7.3) and by equating the two expressions we have:

$$\frac{x}{t_0^2} = \frac{1}{3}\lambda x \quad \Rightarrow \quad \lambda = \frac{3}{t_0^2}. \tag{7.4}$$

This result is in perfect agreement with the observational estimates of $\Lambda = \lambda/c^2$, both in its sign and its value. We can therefore conjecture that the cosmological term is null in PGR gravitational equations (Arcidiacono equations) and that it becomes finite when observations are coordinated with the cosmic time $\tau$ that appears in the customary cosmological models formulated within the context of GR. One must bear in mind that locally (i.e. for $t \approx t_0$) $t$ and $\tau$ coincide, as can be seen

---

useful stratagem. If this were so, the real metric would be the PSR one and the actual gravitational interactions would be included within the interaction vertices where elementary particles are created/destroyed.

directly from (7.1), and that for remote events the observer does not actually measure times but red shift. Thus, the appearance of a cosmological term could be the only truly observable dynamic effect of the difference between $t$ and $\tau$.

To assume the Arcidiacono gravitational equations with $\Lambda = 0$ as fundamental means rejecting the concepts of "vacuum energy" and "false vacuum". $\Lambda$ thus becomes a geometric constant deprived of time evolution, and the possibility of inflationary scenarios during the pre-big bang phase is deprived of any support. These scenarios are radically replaced by the "de Sitter cosmology" outlined above[6].

## 8. COSMOLOGICAL NONLOCALITY AND THE COSMOLOGICAL PRINCIPLE

If the cosmological principle is assumed to be valid then a cosmic time $t$ must exist such that pressure, density and temperature are functions solely of $t$, except for local fluctuations. We note in particular that to assume the existence of a temperature function $T(t)$ means assuming that the Universe is born at $t = 0$ in a state of thermal equilibrium and that it remains in thermal equilibrium during its entire evolution. Naturally, these statements refer to an ideal "cosmic fluid", and condensation phenomena that give rise to the genesis of structures are completely disregarded.

It was suggested above that the singularities associated with the big bang and the big crunch do not consist of two pointevents as in Fridman cosmology, rather of extended $t =$ constant hypersurfaces[7]. Let us consider therefore a single material point coming out of the big bang hypersurface; the portion of the Universe that is visible from this material point is a sphere with radius $ct$, where $t$ is the cosmic time which has passed since the big bang. The point cannot be causally connected to pointevents external to this sphere; nor, therefore, within the limit $t \to 0$, to the pointevents of the $t = 0$ hypersurface. The problem would therefore seem to arise of conciliating the homogeneity of the thermodynamic state in the first instants after the big bang with the locality.

The hypothesis according to which the big bang and the big crunch do not consist of single pointevents in which the entire spacetime collapses but of constant-time surfaces, is currently also being investigated in usual relativistic cosmology, where this argument is technically known as the study of "isotropic singularities" [please see the final section]. However, in the relativistic approach the pre-big bang era must necessarily be inflationary, while in the present approach initial homogeneity can be attributed to nonlocal mechanisms and inflation is not necessary.

Let us consider the two passages illustrated in section 5: they can only occur if $y_0$' and $t$ are defined, in other words if a cosmic time is defined. Thus, the cosmological principle must be verified and, as we have seen, this, in turn, implies that the initial state of the Universe be homogeneous. Thus, only a particular distribution of wavefunctions on the "archaic" 5-sphere (solutions of static 5-dimensional Laplace equations) can activate passages I and II and be explicated in the actually observed material world, in this way becoming the real history of the Universe. The processes induced on the chronotope in this way will, in turn, retroact on the 5-sphere, selecting that distribution as the only one compatible with them. This self-selection of the solution is not a temporal process, because time on the 5-sphere is not yet differentiated from space. The locality condition is therefore not applicable[8].

---

[6] If we assume that $H_0 = 1/t_0$, then the relation $\Omega_\Lambda = \Lambda c^2/3H_0^2 = 1$ follows from (7.4). Therefore an empty PSR Universe with a null cosmological term, for which the relation $H_0 = 1/t_0$ is strictly valid, corresponds in GR to an empty model in which the flatness of space is ensured by the cosmological constant (de Sitter model with expansion). It therefore does not correspond to the Minkowski special relativity chronotope, as one would be led to believe, given the lack of matter and of a cosmological term.

[7] If we assume, as suggested by most recent observations, a Fridman "lambda" model with k = 0, $\lambda > 0$, the big crunch does not exist and the only "future" frontier is the de Sitter horizon at the chronological distance $t_0$. This horizon is not seen by the observer as it is located in his future light cone and he can never reach it.

We can say that, though the equations in question are de Sitter-invariant, the particular solution chosen is not and the breaking of the symmetry thus obtained leads to the cosmological principle and to the appearance of a cosmic time. Before such loss of symmetry, the pure PSR was valid, the Universe was empty and thus appeared to be the same to every observer at every instant (perfect cosmological principle)[9].

## 9. ARCIDIACONO EQUATIONS AND THE COSMOLOGICAL PROBLEM

From a historical point of view, projective relativity was initially developed in the special version (PSR) and only later was extended into the general theory (PGR). Initially [17], gravitation was introduced into PSR through a de Sitter-invariant generalisation of Newtonian theory, along the same line of reasoning which had been expressed in ordinary special relativity by the Poincaré and Nordström theories. The main difference lay in the substitution of ordinary Laplace and D'Alembert operators with their corresponding projective ones[10]. This approach actually shared the same limits as those theories: in these, gravitation is coupled with mass, therefore no gravitational action can exist on bodies whose mass is null. This is not a problem in pre-relativistic mechanics, because bodies whose mass is null do not exist there. But in restricted relativity, as in PSR, bodies with null mass at rest exist. Gravitation acts on such bodies, as is evident in the phenomenon of the curvature of light rays near the Sun's edge.

To explain, on a "Newtonian" basis, the gravitational action on null mass bodies having energy $E$, a gravitational mass defined as $E/c^2$ ought to be introduced for them. There is a drawback to this approach, however; $E/c^2$ is not an invariant, rather it behaves as the fourth component of the quadrivector impulse. In the passage from one inertial reference to another, $E/c^2$ becomes a linear combination of all the components of the quadrimpulse and thus, if we assume: 1) that gravitation is coupled in a given reference frame with $E/c^2$, 2) that this coupling is relativistically invariant, then gravitation must couple with all four components of the quadrimpulse.

The elementary gravitational coupling thus changes an ingoing quadrimpulse into an outgoing one and must therefore be described by a 4x4 matrix. This matrix must give origin to relativistic invariants if the description of the coupling is to be relativistically invariant; it must therefore be a tensor of rank 2 with respect to transformations of the reference frame. Each component of this tensor will be a "gravitational potential", so that the required gravitational theory cannot be based on a single potential, as is the Poincaré theory. The Newton-Poincaré-Nordström theory will be

---

[8] Of course, there is no sense in wondering why the self-selected history of the Universe is precisely the existing one and not another. This would be equivalent to wondering why John Doe is actually John Doe and not someone else. The self-consistency is here relevant instead of a linear causality.

[9] Alternatively, it is possible all the observers coincide at $t = 0$. Due to eq. (7.1), this instant corresponds to $\tau = -\infty$ in the usual "Einsteinian" cosmic time without de Sitter expansion. The interval (-∞, 0] of $\tau$ time should be a sort of pre-big bang era without inflation. In this case, at the "Einsteinian" big bang $\tau = 0$ we should have an already homogeneous Universe having a finite size.

[10] The projective D'Alembertian, translated into ordinary spacetime coordinates, becomes the usual D'Alembertian with the added terms of the $xy/r^2$ type, where $x$, $y$ are distinct spatial coordinates and $r$ is the radius of the Universe. These terms are noticeably different from zero only over cosmological distances from the observation pointevent; locally, the projective D'Alembertian coincides with the usual one. A similar argument is valid for the projective Laplacian.

In homogeneous coordinates the point $x$ coincides with the point $\lambda x$, so that the potential $\varphi(x) = \varphi(\lambda x)$ is a homogeneous 0-degree function, solution of the Laplace projective equation. At great distances from the source mass, the force associated with such a potential deviates from the Newtonian behaviour. Such deviation would be physically detectable only in the case of an enormous source mass, capable of produce great influences at cosmological distances.

recovered in the static limit for massive bodies; in this limit, since only the temporal component of the quadrimpulse (i.e. the mass of the body) survives, only the 00 tensor component will be of relevance. This is what is obtained in general relativity, where in the static case and for weak fields $g_{00}$ is connected to $U$ and the Newton law of gravitation is derived.

It may have been considerations such as these which induced Arcidiacono to formulate the general version of projective relativity (PGR), based on the pentadimensional gravitational equations which now bear his name [3, 9]:

$$R_{AB} - \frac{1}{2} R g_{AB} + \Lambda g_{AB} = \chi T_{AB} \qquad (9.1)$$

In (9.1) we took the permission of adding a cosmological term which, however, based on the considerations developed in section 7, could be null.

In PGR the tensor $g_{AB}$ is generally asymmetrical, as in Einstein's unified field theory; however, the additional components do not have any relationship with the electromagnetic field, as instead is the case in Einstein's theory. The physical interpretation of asymmetry could consist of the fact that if the metric coefficients are measured by means of light ray reflections and time determinations, the measurement carried out through the succession of events ABC may be different from that effected through the succession of events CBA. In this operational sense, therefore, the angle ABC would be different from angle CBA.

Let us introduce the field of quadrics in homogeneous coordinates:

$$\gamma_{AB} \bar{x}^A \bar{x}^B = 0 \ . \qquad (9.2)$$

In the hypotheses discussed above, the cosmological term $\Lambda$ must be null at the level of the (9.1) expressions. It must be an epiphenomenon of the de Sitter $r = ct_0$ radius visible at the GR level, and must vanish in the limit of infinite $r$. In this same limit:

1) the $\gamma_{0i}$s cancel out; this implies the cancelling of the Weyl connection and therefore the absence of a conformal variation of lengths, in accordance with experimental data (remember Einstein's objection to Weyl [18]);

2) $\gamma_{00} - 1$ cancels out; this implies the independence of the gravitational constant of the spacetime coordinates, in accordance with observational results;

3) the asymmetrical part of $g_{AB}$ vanishes.

Thus, the Einstein equations are simply "bordered" by the terms $0i$, $00$; naturally, it is possible (and probable) that conditions 1)-3) are verified also with finite $r$.

As we can see, PGR has a substantially greater wealth of solutions and particular cases than does ordinary GR. The mathematical problem associated with the cosmological issues discussed in this article is therefore considerably complex and still awaits to be solved.

Through a simple argument developed in Newtonian approximation [6] it can be shown that the (9.1) expressions admit solutions which satisfy the cosmological principle and which connect with the Fridman ordinary relativistic cosmology solutions in the limit $r \to \infty$. Such solutions could constitute the asymptotical tail of solutions corresponding to isotropic singularities, and this is the hypothesis formulated here for the big bang and the big crunch. The reason why we have been led to consider it probable is that in PSR the observer sees - as the only singularities - the past and future de Sitter horizon placed at the chronological distance $t_0$ from him. For the relativity principle, this is true for every observer in uniform rectilinear motion with respect to him and this implies that the singularity in question cannot be constituted by two isolated pointevents but, indeed, by two boundaries. One might believe that in PGR the big bang singularity seen by a chronologically very remote observer (for example, a current observer) appears "crushed" or "compressed" onto the past de Sitter horizon. This would be a "soft" introduction to the PGR effects on the PSR basic scenario, which must be continuously approached by PGR in the limit of indefinitely low densities of matter;

the big bang would thus acquire a spatial extension. In order to have, at the big bang, a finite mass and infinite density, the metric must become singular, so that the measures of distance and volume cancel out. This is equivalent to saying that the big bang must be an isotropic singularity.

By adopting the definition developed by Anguige and Tod [19] within the context of general relativity, a spacetime $M'$ with metric $g_{ik}'$ is said to admit an isotropic singularity if a variety with border $M$ included within $M'$, a regular Lorentz metric $g_{ik}$ on $M$ and a function $\Omega$ defined on $M$ exist such that

$$g_{ik}' = \Omega^2 g_{ik} \qquad \text{for } \Omega > 0 \qquad (9.3)$$

$$\Omega \to 0 \text{ on } \Sigma$$

as $\Sigma$ is a smooth spacelike hypersurface in $M$, called *singularity surface*. Isotropic singularities form a singularity class having a finite Weyl tensor. The study of cosmological models with this type of singularities began recently within the context of usual relativistic cosmology, while relevant literature is completely lacking in PGR.

The authors hope that the physical context hypothesised in this article can stimulate the mathematical work necessary to identify possible solutions of eqs. (9.1) having the characteristics indicated.